# Data Protection by Design for cybersecurity systems in a Smart Home environment

Olga Gkotsopoulou*, Elisavet Charalambous†, Konstantinos Limniotis‡, Paul Quinn*, Dimitris Kavallieros§, Gohar Sargsyan**, Stavros Shiaeles††, Nicholas Kolokotronis‡

* Vrije Universiteit Brussel, Belgium. Email: Olga.Gkotsopoulou@vub.be; Paul.Quinn@vub.be
† ADITESS Ltd., Cyprus. Email: lc@aditess.com
‡ University of Peloponnese, Greece. Email: klimn@uop.gr; nkolok@uop.gr
§ Center for Security Studies (KEMEA), Greece. Email: d.kavallieros@kemea-research.gr
** CGI, The Netherlands. Email: gohar.sargsyan@cgi.com
†† University of Plymouth, UK. Email: stavros.shiaeles@plymouth.ac.uk

*Abstract*—The present paper deals with the elucidation and implementation of the Data Protection by Design (DPbD) principle as recently introduced in the European Union data protection law, specifically with regards to cybersecurity systems in a Smart Home environment, both from a legal and a technical perspective. Starting point constitutes the research conducted in the Cyber-Trust project, which endeavours the development of an innovative and customisable cybersecurity platform for cyber-threat intelligence gathering, detection and mitigation within the Internet of Things ecosystem. During the course of the paper, the requirements of DPbD with regards to the conceptualisation, design and actual development of the system are presented as prescribed in law. These requirements are then translated into technical solutions, as envisaged in the Cyber-Trust system. For trade-offs are not foreign to the DPbD context, technical limitations and legal challenges are also discussed in this interdisciplinary dialogue.

## I. Introduction

The interconnectivity of consumer devices in a Smart Home environment - from smart TVs, lightbulbs and coffee makers to smart grids and smart security systems - upsurges constantly. An increase in connectivity may entail more risks for the end-users of those devices in combination with the rise in cyberthreats and cybercrime incidents. Often, these products come with different levels of embedded security features depending on the country of their origin or sale and until recently, due to the absence of strong monetary incentives for manufacturers to fabricate more secure devices [1]. To that can also be added the lack of standardisation in the communication protocols and practices followed in the operation of Internet of Things (IoT) devices.

To address this situation, legislators in European Union Member States and other third countries request the integration of minimum built-in security features which would protect both the integrity of the connected devices and the data on them. On the other hand, tougher obligations are imposed upon service providers to ensure security and integrity of their services with the deployment of proper cybersecurity systems. Amid these circumstances, the Cyber-Trust project offers a holistic cybersecurity solution, which develops around three pillars: proactive cyberthreat intelligence technologies, machine learning based detection and mitigation tools as well as Distributed Ledger Technologies (DLT). Based on those parameters, the Cyber-Trust project aims to ensure: a. the maintenance of the integrity of systems used by various service providers as well as by individual end-users, safeguarding selected devices and identifying vulnerable or improperly manufactured IoT appliances that could endanger all networked apparatus; and b. the amplification of cooperation between service providers and Law Enforcement Agencies (LEAs) with regards to the preservation and transfer of electronic evidence, when a cyberattack takes place and is mitigated.

Given that service providers as data controllers are accountable for their cooperation with proper data processors, such accountability could also exceed in their obligation to select software and IT products which are data protection compliant. In this paper, our research is focused on the Cyber-Trust solution from a Data Protection by Design perspective, by adopting an interdisciplinary approach and using as a starting point previous and on-going research conducted throughout the project. In order to examine how a cybersecurity product can embrace the Data Protection by Design principle as set out in Article 25 of the General Data Protection Regulation, we structure the paper upon two main drives. After discussing what the legal obligations are (first drive), we will see how those can be translated into technical solutions (second drive). Last, the remaining challenges in the form of limitations during the research phase as well as post-research will be analysed both from a legal and a technological point of view.

## II. A brief description of the Cyber-Trust project

The Cyber-Trust project aims to develop an innovative cyber-threat intelligence gathering, detection, and mitigation platform to tackle the grand challenges towards securing the ecosystem of IoT devices. The security problems arising from the flawed design of legacy hardware and embedded devices allows cyber-criminals to easily compromise them and launch large-scale attacks toward critical cyber-infrastructures. Cyber-Trust is designed especially for application in the Smart Home and mobile devices domains where a wide range of functionalities of varying complexity and implementation strategies are implemented.

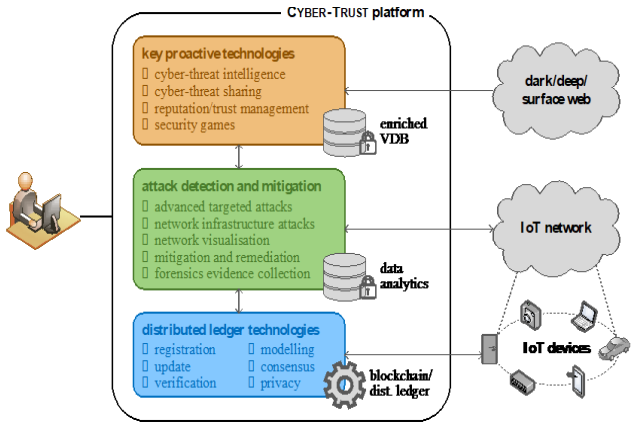

Figure 1: High-level overview of the proposed cyber-security platform

The necessity to monitor end-user IoT devices results in great challenges in preserving the privacy of end-users whilst gathering enough information for efficient and reliable detection of cyber-attacks. One of the primary goals of Cyber-Trust is to facilitate information sharing between various key stakeholders and therefore, seeks deployment at a large scale. Due to this, Cyber-Trust provides end-users flexibility in the mode of operation, meaning that the latter may select which of their devices will be protected (partial or full provision) and whether registered devices will be in for passive or active monitoring. Additionally, to allow further flexibility, Cyber-Trust is designed to support multi-tier registration, at organisation, user and device level for full control.

Cyber-Trust is designed to be self-evolving with feeds from external threat intelligence sources, knowledge bases as well as anonymised end-user information for the early detection and response to persistent cyber-attacks and newly discovered/exploited vulnerabilities. Cyber-Trust implements a profiling service that correlates, and matches gathered external intelligence with internally acquired data. The core system gathers information from multiple sources: published vulnerabilities, latest patches, firmware releases with the latest security updates, manufacturer use guidelines, user provided information and runtime information. User provided information involves the data and setup provided by the end-user since registration, while runtime information is collected when the end-user enables active monitoring for each device individually; enabling active monitoring on a gateway results in monitoring network transactions from all connected devices.

III. MAJOR CHALLENGES REGARDING DATA PROTECTION IN A CYBERSECURITY CONTEXT IN THE SMART HOME ENVIRONMENT

It is well-known that, generally, several data protection risks appear in the context of Smart Home. More precisely, IoT consumer devices process several types of personal data, such as the end-users´ location or their daily habits, which are transmitted to central servers or shared with other devices or third parties. This, in turn, raises several concerns in terms of the end-users' right to personal data protection, since there is a possibility of processing these data for other purposes (e.g. for profiling the users via fully-automated means), without obtaining their consent or having other legal ground, for instance. Clearly, such risks necessitate the adoption of proper data protection mechanisms.

Although appropriate security measures suffice to also alleviate some data protection concerns – for example, regularly updating device firmware is an important step towards thwarting several attacks which could yield in compromising personal data confidentiality – there are often trade-offs between them. This is mainly attributed to the fact that, towards enhancing security, there is probably a need to massively collect several types of data to be analysed, which in turn may pose personal data protection at risk (e.g. in terms of respecting the principle of data minimisation). Such trade-offs will be thoroughly examined in the context of Cyber-Trust, since gathering and appropriately handling information from several sources constitutes, as previously stated, a crucial building element for the whole system.

Within the aim to frame the key recommendations for the conceptualisation, design and actual development phase of the Cyber-Trust project, the technical partners shared their main preliminary concerns identified in a technical level regarding the data protection (and privacy) requirements as well as respective potential risks. The difficulty in pre-assessing whether personal data will be processed in each specific use-case, as well as the likelihood of a personal data breach, including the incidental disclosure of information of sensitive character about an end-user´s device, were identified as the primary challenges to be tackled. Other concerns referred to the use of tools, such as data filtering techniques in order to achieve data minimisation, the implications deriving from the use of Deep Packet Inspection (DPI), the case-by-case assessment of proportionality and finally, the handling of material that may contain electronic evidence, including measures to avoid unauthorised access by third parties. This preliminary assessment performed by the technical partners at the conceptualisation phase is used as a ground which the Data Protection Impact Assessment (DPIA) envisaged for a later stage, will be shaped upon.

IV. THE PRINCIPLE OF DATA PROTECTION BY DESIGN IN THE CYBERSECURITY CONTEXT

A. General considerations

Since 2010 the Article 29 Working Party argues that "[d]ata protection must move from 'theory to practice'. Legal requirements must be translated into real data protection measures", [2] in order to tackle risks arising from innovative technologies and protect end-users of Information and Communications Technologies (ICT) services and products with limited or average knowledge and skills [3]. Data Protection by Design constitutes a critical combination of technology and law [4] and has always been part of data protection legislation as one of the most intense forms of cooperation between different disciplines. Even though DPbD is often interchangeably referred to as Privacy by Design since the second term appeared earlier and many main notions underlying both principles, it becomes clearer that these two are different concepts questing for separate assessment [3]. Despite its undoubted significance, it was only in 2016 that DPbD was expressly stated as a legal requirement in the European Union law in the General Data

Protection Regulation (GDPR) [5] along with Data Protection by Default in Article 25.

By implementing appropriate technological and organisational measures, data controllers are called to consider data protection and privacy issues upfront: not only at the time of the processing itself but also at the time of the determination of the means for the processing. before the data controllers are engaged with any data processing operation or the development of a new product. Data controllers are expected to take into account the nature, scope and context of the processing. Although DPbD turns the focus on accountability, it does not entail only a positive obligation of the data controller to take action or implement a measure. Foremost, it is an obligation from result, meaning that best efforts are not sufficient; concerned entities must achieve DPbD and be able to prove that they did so [3]. Even though it is not explicitly stated in the provision, data processors, when processing personal data on behalf of the data controller, will also have to implement relevant measures in order to contribute to data controllers´ compliance obligations, as provided in Articles 4(8) and 28 GDPR.

Moreover, the DPbD principle is thought to extend to hardware and software providers [3]. More precisely, it should be pointed out that, according to Recital 78 GDPR, the producers of the products, services and applications that are based on the processing of personal data or process personal data should be encouraged to take into account the right to data protection when developing and designing such products, services and applications, as well as to make sure that controllers and processors are able to fulfil their data protection obligations, by taking into consideration the state of the art ("due regard"); this is the only reference in the GDPR to stakeholders others than the data controllers or data processors, thus further illustrating the importance of the DPbD principle.

To sum up, the entities directly involved in the implementation of the DPbD principle are the data controllers, the data processors and the producers. Albeit, it is noted that end-users of a product or operators of a system may be able to alter its configuration; in this paper the term end-users refers to individual customers whereas operators refer to employees who work with the system having specific access rights and process personal data representing the data controller or data processor [17].

National supervisory authorities [7], legal scholars [3] as well as European Union independent institutions, bodies and agencies, such as the European Data Protection Supervisor [8] and the European Network and Information Security Agency [9] have attempted to provide sets of practical requirements. These guidelines aim to assist data controllers when choosing among variants of existing software, or producers who engage in the creation and development of in-house new technological tools and systems intended to be used for data processing operations, to comply with this new legal requirement and stay clear from sanctions and administrative fines. Although these guidelines do not refer explicitly to cybersecurity products, it is implied that as long as personal data processing occurs, DPbD, as provided in GDPR, is applicable, unless the product or the system is exclusively deployed, for instance, in a law enforcement context where specific legal instruments could apply as *lex specialis* [10].

In general, what could be considered as an appropriate organisational and technical measure must be interpreted dynamically depending on the context of the specific application and the emerging risks. Article 25(1) and Recital 78 GDPR include a non-exhaustive list of measures that can be taken, with pseudonymisation being a prominent example. However, it should be highlighted that technical or organisational measures are not standalone; a technical measure may only be effective with the parallel implementation of organisational measures, and the same applies *vice-versa*.

As seen above, GDPR adopts a risk-based approach. Thus, the concerned entities have to first explore the state of the art concerning processing means, by carrying out extensive market research and remaining on top of the most recent technological developments in standardisation and cybersecurity [3]. The maturity of those solutions, as well as the cost of implementation, must be taken into account. Only the "Best Available Techniques" (BAT) are in principle to be selected [8].

### B. The requirements of DPbD in a nutshell

The requirements of DPbD - even though no list can be exhaustive, as this was not the intention of the co-legislators - must address in the most effective manner the data protection principles, the data subjects´ rights and provide safeguards to accommodate any other requirements set in GDPR [3]. Like the DPIA, which in fact constitutes a considerable part of DPbD [6], it can be regarded as another tool applying similar methodology, aiming to enhance compliance with the fundamental principles and requirements set out in the law and boosting data subjects´ protection.

The determination of data protection goals could be served through various privacy engineering methodologies depending on the adopted approach [8]. For example, some methodologies invest in a risk-management approach, whereas others prefer the creation of design patterns [8]. Nevertheless, most of those methodologies are driven by the six main protection goals which are usually summarised as: confidentiality, integrity, availability, unlikability, transparency and intervenability [21]. In the US context, the last three terms would be replaced by predictability, manageability and disassociability [8]. Those six goals will be discussed in the next three subchapters.

#### 1) During the conceptualisation stage

First of all, as mentioned above, consideration of data protection issues has to take place already at the early stages of conceptualisation and design and not only during the implementation of systems and business practice [7]. This is the core of the DPbD principle, i.e. data protection must be seen as an essential and indispensable component of every processing system from the very start. Second, preliminary consideration of the envisaged data processing activities at the conceptualisation stage should include a number of steps:

the concerned entity shall conceptualise the data processing by clarifying whether the tool or system will be intended to proceed personal data and which types of personal data. Furthermore, it should determine whether the use of the system falls under any of the exceptions where special types of personal data can be processed.

A useful tool to that end could be the creation of data flow diagrams via data flow mapping of the different data processing activities [8]. Those diagrams will be re-assessed in the design and actual development stage as well as in the relevant Data Protection Impact Assessment, wherever necessary.

Next, the concerned entity shall identify (possible) data controllers and processors or subcontractors and establish - if necessary, in the particular context - contractual relations. When the creator of the system is either the data controller or data processor, the concerned entity should already identify a legal basis and a purpose for the processing, as well as with respect to storage limitation, the time the data will be kept and processed, without excluding the possibility for the need to set criteria and plan ahead for automatic erasure. Quite early in time, the concerned entity shall also assess the need or the possibility for cross-border transfers, since transfers would require specific conditions upon which they can be considered lawful.

Equally imperative questions to assess at this stage are [7]: how transparency of the activities is planned to be achieved. For instance, transparency could be enhanced with the introduction of an easily accessible data protection and privacy settings dashboard, which would give the end-users the possibility to easily exercise their rights; second, since every concept would have to be assessed separately, the context of the data processing must be defined, since the use of the product by LEAs or by individual consumers would result in different legal requirements, as mentioned earlier in the text.

Last but not least, the concerned entities would have to determine whether special requirements apply in their sector, if there are codes of conduct or certification schemes that they could benefit from, as well as whether there exist guidelines or decisions issued by the respective national supervisory authorities or national courts, specifically addressing same use cases. In particular, Article 25(3) GDPR underlines that adherence to certification schemes as provided in Article 42 can be considered as a way of demonstrating compliance.

*2) During the design and actual development*

In the previous paragraphs, we briefly listed the requirements regarding the preliminary assessment of the DPbD in the conceptualisation phase. Requirements pertaining to data protection principles to be taken into account during the design and actual development phase include the following:

**Lawfulness, fairness, and transparency**: processing of personal data is only lawful if one of the following holds: the data subject has given her/his consent; the processing is necessary for the performance of a written agreement or contract with the data subject; for compliance of a legal obligation to protect vital interests of a natural person; to perform a task carried out in the public interest or in the exercise of official authority; and last but not least, for a legitimate interest pursued by the data controller. Specifically, when the legal ground is consent, the latter must be - according to Art. 4(11) of the GDPR - freely given, specific, informed and unambiguous (i.e. by a statement or by a clear affirmative action), given that consent cannot be implied. Although consent is not required to be given in written form, provision in written (or in electronic) form is recommended due to the accountability requirement of the data controller. Moreover, users must be able to withdraw consent at any time without negative consequences and should be facilitated to do so, if they wish.

It should be made possible that clear and comprehensible information is provided to the data subject regarding the purpose of the processing of personal data, the legal basis and the recipients of the information [19]. This can be fulfilled by designing privacy-friendly user interfaces which avoid dark patterns and reduced usability [20]. However, the provision of comprehensible information presupposes analytic documentation and that the system functionalities are thoroughly understood by both the technical and legal partners involved in the development of the solution. For instance, consent forms bundled with terms and conditions for the use of a service do not constitute a valid practice. If the last ground of legitimate interest is invoked, then a balancing between the interests of the data controller and the freedoms and rights of the data subjects should take place. In such a case, to ensure a proper balance, appropriate safeguards are generally needed to be in place.

Accordingly, a strong data protection default should be established, along with data protection settings which would allow the end-user to effectively control without any burden his/her preferences. The influence of Data Protection by Default to the system design will be assessed to a dedicated subchapter below. The processing of personal data must be transparent, meaning that the system and its components must be designed in a way that relevant aspects of personal data processing are known to the data subjects, enabling them to make informed decisions and exercise their rights. Furthermore, the tool must ensure that other rights, such as privacy, freedom of expression and absence of discrimination are likewise safeguarded [7].

As part of transparency during the design and development phase could be also considered the keeping of records or relevant documentation, that show how the aforementioned requirements have been implemented, which vulnerabilities have been discovered and which methodologies have been used in respect of data protection. This documentation could also contribute to the provision of clear and unambiguous information to the data subject as seen earlier [7].

**Purpose and storage limitation**: personal data shall only be collected for specified, explicit and legitimate purposes and shall not be further processed in a manner which is incompatible with the original purposes. Moreover, the data shall be kept for no longer than it is strictly necessary for the purposes pursued. Personal data should be anonymised or deleted when the purpose of processing is fulfilled. As mentioned above, data flow mapping may be necessary in order to keep track of those obligations.

**Data minimisation**: only adequate, relevant and limited to what is necessary for the purpose pursued data should be processed. It is self-evident that if a tool can work equally satisfactorily with the use of non-personal data, then personal data shall not be in principle processed.

**Accuracy**: mechanisms shall be envisaged to keep all personal data accurate and up-to-date, whereas ensuring that incorrect data can be deleted or rectified as quick as possible.

**Integrity and confidentiality**: appropriate security measures shall be implemented to ensure the confidentiality, integrity and availability of the data, such as vulnerability-reducing mechanisms.

In addition, the tool must enable the data subject to exercise his/her rights, satisfying the goal of intervenability [7]. These rights are: the right to access their personal data, information about the processing, and other rights; the right to rectify their personal data; the right to delete their personal data, if applicable; the right to restriction of processing of their personal data, if the conditions for restriction apply; the right to data portability for their personal data, if the processing is based on consent or agreement and is carried out by automated means; the right to object against the processing of their personal data, if applicable in the specific case; rights relating to automated individual decision-making, including profiling that may have legal consequences, or similarly significant effect for the person concerned.

Note though that, in relation with the data minimisation principle, if the personal data processed by a controller do not allow for identifying the data subject, the data controller should not be obliged to acquire additional information in order to identify the data subject for the sole purpose of complying with the above provisions (see Recital 57 of the GDPR).

*3) Impact assessment, testing and post-research*

All the aforementioned if´s are assessed as well in the DPIA. Trade-offs that may be encountered depending on the circumstances cannot be necessarily solved but can be explored with respect to data subjects' rights and interdisciplinary research, aiming to limit mismatches between the law and the practice [11].

Moreover, the creators of the system must allocate sufficient time for multilayer testing, in order to verify that the planned DPbD measures were actually implemented and they are indeed functional. The Norwegian Data Protection Authority recommends different tests to be performed before the release of the system, such as data protection requirements testing to ensure the integration of all the envisaged measures in the conceptualisation and design phase and dynamic testing in order to secure functionality in case of system failures and restrain risks when different user permissions are enforced [7].

Since cybersecurity is a constantly evolving field, and a cybersecurity system as envisaged by the Cyber-Trust project is rather dynamic, Data Protection by Design requirements would also cover the addition of new features and any modifications in the technical measures or organisational solutions as well as the maintenance of the system [7], in the case the Cyber-Trust final product is commercially launched.

*C. Data Protection by Default and System Design*

DPbD and Data Protection by Default are two concepts, which are closely inter-related, not only because they constitute part of the same provision but primarily because default functionalities concerning data protection are introduced during the design of a system and can be more or less extensive, given the causes pursued [17]. The effect such a default setting may have to data subjects' rights becomes obvious by the fact that data protection settings may remain unchanged for the whole duration of the use of a product, even if data subjects are given a choice to allow or deny the processing of their personal data in a broader way [18]. In other words, the data subject, as well as the operator of the system, may not wish to interfere with those settings.

As simply put by the European Data Protection Supervisor "[t]he idea behind the principle [of Data Protection by Default] is that privacy intrusive features of a certain product or service are initially limited to what is necessary for the simple use of it" [18]. Since default settings can influence the overall architecture of a system, they are, therefore, in the heart of the DPbD and should be understood as not requiring active behaviour from the end-use side [8]. For instance, privacy nudging mechanisms on the one hand and facilitated and uncomplicated opt-out processes on the other fall under both the scope of DPbD and Data Protection by Default [17].

The strong ties between the two principles implement that the one cannot be taken into consideration without the other and create an efficient net of all-encompassing solutions which satisfy both requirements.

V. TRANSLATION OF THE DPbD RECOMMENDATIONS INTO TECHNICAL SOLUTIONS

Data in Cyber-Trust reside on four layers: end-user, service provider, external knowledge bases and LEAs. Depending on the user configuration and mode of operation the two latter layers may be inactive. This is the case when the device owner registers a device for passive monitoring which entitles the reception of notifications and information regarding firmware releases and emerging vulnerabilities related to their device. Regardless of operation, information residing within the Cyber-Trust needs to remain secure in all of its stages: acquisition, communication, storage, dissemination. Additionally, authority management is a key concept of the Cyber-Trust DLT regarding the sensitivity of the data stored inside it. Indeed, the forensic evidence stored in the DLT and private information relative to an attack are critical.

The Smart Device Agents (SDA) and Smart Gateway Agents (SGA) are the two Cyber-Trust components responsible for the acquisition of information from end-user IoT devices and gateways respectively and represent the links with the Cyber-Trust core components hosted on the service provider layer. Monitoring of the end-user's gateway is by default inactive as it enables active monitoring for all connected devices and the need to transfer exchanged traffic to the Cyber-Trust backend for DPI; the end-users may enable/disable this option through their profile at any time only after they have clearly consented to the processing of their personal data. In contrast to SGA, SDA operates in a more restrictive manner as its purpose is to receive information regarding new vulnerabilities and modes of operation from the Profiling Service and to communicate back in the occurrence of a suspicious event.

As a form of data minimisation and a measure of gathering only data that serve legitimate purposes, the SDA exhibits intelligence and performs real-time monitoring. Different flavours of SDA will be implemented within the framework of Cyber-Trust as a range of smart devices need to be accommodated with mainly two modes of operation: one being for continuous/ real-time operation and the latter for ad-hoc operation when the circumstances call for it.

The SDA is responsible primarily for the monitoring of device's usage, critical files, security status (patching status, firmware integrity, vulnerability risk) as well as suspicious network transactions, and secondly for the application of mitigation policies and remediation actions after the detection of an attack or threat that could endanger the integrity and operation of the monitored device. Due to its intended operation, the SDA is designed to check whether the hosting device performs as intended by its manufacturer, ensures that critical OS files are uncompromised and that only secure means of communication are used. Data regularly synched with the Profiling Service involve information regarding runtime processes and used hardware resources. Only in the case of identified suspicious traffic and activity, network packages are signed by SDA and communicated with the Cyber-Trust Cyber Defence service for further investigation; metadata regarding this activity are registered on the DLT for future reference.

Data from the SDA and SGA are communicated to the Profiling Service (PS), responsible for the storage and management of the Cyber-Trust generated and acquired data. In particular, two separate access control layers are supported: one for defining architectural policies and one for controlling runtime operations for matching use preferences. Amongst others, the system implements end-user anonymisation or pseudonymisation (if desired), data ageing and access mechanisms, targeted communication from and towards the LEAs, as well as dynamic progress control and accountability mechanisms.

The PS acts as a multi-faceted centralised data management system enforcing security on the data level, adhering high modularity and horizontal scalability. The PS ensures authorised and justified access to maintained data, including multimedia and binary content supporting the implementation of advanced searching capabilities. Data integrity of the original information is preserved with data hashes and digital signatures for any embedded multimedia object. The original data along with integrity preserving metrics are stored in a separate repository in which no operator of the system may perform changes, while data concerning forensic use and of value to LEAs are stored, encrypted, in a separate repository where only authorised LEA users are allowed access.

Data ageing attributes are attached to the object ensuring data retention only for the necessary and predefined period of time. The renewal of an item's date of expiration is possible and needs to be initiated from an operator with the appropriate access key (separated from other functional permissions). The PS also implements the mechanisms for maintaining extensive system logs and auditing. All incoming and outgoing actions/requests are logged and retained to the system in the form of a record; only visible to operators with appropriate permissions.

Communication is achieved through Rest API over a security channel (SSL/TLS encryption over HTTP), for the insertion/addition, retrieval, update, and searching of data. As the PS implements extensive mechanisms for data and operators access rights, the provision of access tokens is also necessary for internally preserving security on preserved content. Internal security mechanisms protect the end-user against data protection and privacy concerns as well as the un-lawful access on data by system operators.

The Cyber-Trust proposes a sound solution for the monitoring and safeguarding of end-user devices, combining a range of technologies with a multi-tier and modular architecture. Limitations and restrictions for its operation are subject to the end-users' setup and compatibility between monitored devices. The method of architecture design in Cyber-Trust is the proven Risk- and Cost-Driven Architecture (RCDA) based on advantages versus other approaches [15]. The advantage of applying this method is that it supports architectural and design decision making throughout the whole design process started from design. Concerns and decisions are weighed throughout the design process, and stakeholders' requirements are constantly validated against the design. The design process is iterative to ensure high quality results. The fact that RCDA is a recognised method in the Open Group Certified Architect program, it is an extra advantage for the project and consortium partners to promote openness and collaboration on the most efficient way of shaping the design and architecture.

## VI. REMAINING CHALLENGES

### A. Technical limitations

The heterogeneity in the design, capabilities and purpose of IoT devices impose limitations that the Cyber-Trust platform is aiding to bring at the owner's knowledge. Smart devices are generally classified in resource constrained devices and high capacity devices. The limitation of resources leads to a lack of security mechanisms especially when the IoT device is set-up for ease of use. Moreover, traditional security practices may not be applicable, installation of software updates may be hard or impossible because of the device's complexity or the product being discontinued by the manufacturer. For resource constrained devices with adequate API support, an additional SDA agent may be deployed as middleware between the devices and the SGA.

Additional limitations are imposed by the connectivity supported by the IoT devices as the latter may communicate through channels that the SDA and SGA do not support (i.e. Bluetooth, ZigBee); in these cases, monitoring of devices can be performed at the hub level (if one exists), potentially hindering the system's effectiveness in identifying malicious activity. As a result, the satisfactory provision of Cyber-Trust functionalities depends on the capabilities of the end devices, and this is a limitation that no system may overcome.

Even though the SDA is designed to fit a range of IoT devices, the development and maintenance of different versions of the SDA to capture applicability to enough smart devices imposes a great challenge especially considering that drastically different approaches are used by IoT manufacturers for the production of similar devices.

*B. Legal Compliance*

Perhaps the most important challenge with regards to legal compliance is to bring legal and technical understanding under common terms. Data protection – as well as privacy - in legal terms is not a synonym of security in technical terms. Data security constitutes only one of the obligations for data controllers (Article 32 GDPR) and one single part of a complex system of checks and balances established under the data protection law. Often, the notion of data protection is wrongfully mistaken for its etymological meaning by software engineers and developers, in other words, as the application of strong technological measures to safeguard the data [11]. Their perception does not usually extend to the "if's" and "how's" of the collection or other processing happening to the data as long as strong security safeguards have been implemented. To those misunderstandings also contributes the "vague principles" argument [5], often attributed to the Privacy by Design notion, the non-legally binding "predecessor" of DPbD concerning privacy. Nevertheless, it is argued that the data protection principles and data subjects rights constitute a more robust ground for implementation of specific requirements as discussed by a number of legal scholars [11]. Effective implementation of those requirements would prove by result that DPbD was indeed achieved, as mentioned above.

Tthe fact that the list of organisational and technical measures, as provided in GDPR, is only indicative, could aggrade the aforementioned challenge, when data controllers and producers are called to assess and select the measures. Despite the existence of guidelines, the use of international standards national and accreditation systems and the efforts of the European Standardisation Organisations [8], what misses at the moment is a pan-European mechanism to establish that the requirements and those auxiliary recommendations at the conceptualisation and design phase, were in fact taken into consideration and implemented into the final product [12]. The EU Cybersecurity Act [16] currently under approval by the European Parliament may introduce a certification scheme for ICT products, which hopefully could lead to further legal clarity and the deployment of common standards. Depending on the concluding outcome, the co-legislators may call for additional requirements and shed light to the implementation of appropriate organisational and technical measures "in an effective manner", as appears in Article 25(1) GDPR. Moreover, in the course of time other standardisation procedures in relation to Articles 42 and 43 of GDPR initiated by Member States and including "requirements for manufacturers and/or service providers" to implement the principles "applicable to all business sectors, including the security Industry" could create a baseline for the state of the art and enhance country-wide best practices, as a point to start with [8].

Another misconception that has to be explicitly addressed, in order to enhance the common understanding among the different involved parties, is that data protection compliant does not mean privacy compliant. A privacy infringement does not presuppose a data protection infringement. The final product most likely will also have to meet the privacy requirements, which appeals to further consideration from both legal and technical partners; for instance, when conducting the overall Impact Assessment before its deployment, and therefore Privacy by Design shall also be a factor to be reflected upon [13].

In addition, a sophisticated and complex system like Cyber-Trust consists of both in-house solutions and third parties' tools, depending on the available software kits. Concerning the latter, assessing each one of them, as required by the DPbD, *a posteriori* (meaning after they have already been created by someone else and in a different context) could be rather a time- and resources- consuming or even impossible, provided the available information and documentation. Even though there are applications which help with the evaluation of such software, this would still imply dependence on third parties' evaluation criteria. As for the in-house solutions, they would also require component-by-component and case-by-case assessment. In other words, every single tool/system has to be assessed separately in the specific context it is intended to be used and for each envisaged use, taking into account interdependencies and function creeps. Such an assessment would require a very high degree of technical comprehension from both the technical and the legal partners and could be also rather time-consuming and resources- demanding task.

Further, it should be pointed out that anonymising personal data, which could possibly be considered as an appropriate privacy enhancing technology, needs to be very carefully considered. Although there exist several anonymisation techniques, none of them should be considered as panacea; as Recital 26 of the GDPR states, "to determine whether a natural person is identifiable, account should be taken of all the means reasonably likely to be used by anyone to identify the natural person directly or indirectly, taking into account all objective factors". In other words, before characterising data as anonymous and therefore, as non-personal, we need to cautiously answer the question whether it is impossible for any party – including the data controller - to identify from these data any individual [14]. There are many famous examples in the literature with regards to "bad" anonymisations which allowed for re-identifying some of the end-users in an

"anonymised" set. Hence, in practice, it is highly questionable whether, in the case of the Cyber-Trust system as in any other system, real anonymisation can be achieved right now with the existing technical means, given also the 2014 Opinion of the Article 29 Working Party on anonymisation techniques [14]. Therefore, anonymisation techniques should be considered in the Cyber-Trust framework and other similar cases as mechanisms to promote data protection rather than to actually render data as anonymous.

Similarly, great challenges also occur in the process of determining the proper pseudonymisation methods. According to Recital 28 of the GDPR, "the application of pseudonymisation to personal data can reduce the risks to the data subjects concerned and help controllers and processors to meet their data protection obligations. The explicit introduction of 'pseudonymisation' in this Regulation is not intended to preclude any other measures of data protection". Pseudonymisation generally refers to hiding the end-users' identities, whereas in the case of the Cyber-Trust system for a Smart Home Environment, this could possibly necessitate the hiding of devices' identities, too. Choosing, though, a proper pseudonymisation approach constitutes a non-trivial task. To this end, all the relevant data protection risks should be carefully considered, whilst the proper overall operation of the Cyber-Trust system so as to fulfil its primary security goals should be ensured.

Finally, most, if not all, technical measures have to be accompanied by organisational measures. Policies have to be put in place to enable the rectification and deletion of personal data, to keep some areas restricted to only authorised individuals and other areas public to registered entities as well as to provide information on data subjects' rights. Since the Cyber-Trust system – as is the case with other cybersecurity solutions – aims to secure the protected systems of the service providers from cyberattacks, facilitate the preservation of material which may contain electronic evidence and offer a safe and integral way of transfer of the latter from service providers to LEAs, wherever the right legal authorisation exists, such a Daedalian system would require a preliminary assessment of the implementation of organisational measures as well. Such preliminary assessment would have to, later on, be re-assessed by the actual data controller and adapted in the needs of the specific data processing.

## VII. CYBER-TRUST PILOTING

The integrated Cyber-Trust prototype solution will be released in early 2020 to undergo two piloting cycles, capturing the needs of its major stakeholders. The first piloting cycle is planned to last four months and will focus on the platform's capability in the detection and mitigation of cyber-attacks and the solution's impact when deployed in a large scale by an Internet Service Provider (ISP). After the end of this phase, the integrated platform will be refined with further fixes and updates, as these emerge from the needs of the ISP. The final version of the platform will then be tested to its full capabilities, for a two-month period, by also involving the second key Cyber-Trust stakeholder, active LEAs (Cybercrime investigators and Digital Forensic examiners). Cyber-Trust committed to the delivery of advanced cyber intelligence tools to the open-source community will publish its final product publicly on GitHub.

## VIII. CONCLUSIONS AND FUTURE WORK

The principle of Data Protection by Design as a binding requirement for data controllers was introduced in the EU Member States law with the General Data Protection Regulation. Deriving from the principle of accountability, such a requirement according to Recital 78 GDPR seems to extend also to manufacturers and providers of software and other IT systems. The DPbD refers to the implementation of technical and organisational measures which aim to ensure the data protection principles and safeguard data subjects' rights already in the very first stages of the creation of a new product or a system, namely the conceptualisation, design and actual development.

The discussion was based on the research conducted and the conclusions drawn within the Cyber-Trust project and is pertaining both to the research phase and the post-research phase concerning the launch of the final product. So far, the Cyber-Trust helped us to study how the Data Protection by Design requirements can be integrated into a cybersecurity system, given the trade-offs established by technical limitations and legal challenges. However, the finest details will be tuned in the forthcoming pilots and will be assessed in the two Data Protection Impact Assessments that will follow, requiring further interdisciplinary research and cooperation.

ACKNOWLEDGMENT

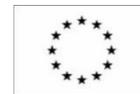

This project has received funding from the European Union's Horizon 2020 research and innovation programme under grant agreement No 786698. The work reflects only the authors' view, and the Agency is not responsible for any use that may be made of the information it contains.